# EVOLUTION OF OXYGEN ISOTOPIC COMPOSITION IN THE INNER SOLAR NEBULA

ALEXANDER N. KROT[1*], IAN D. HUTCHEON[2], HISAYOSHI YURIMOTO[3], JEFFREY N. CUZZI[4], KEVIN D. MCKEEGAN[5], EDWARD R. D. SCOTT[1], GUY LIBOUREL[6,7], MARC CHAUSSIDON[6], JEROME ALÉON[6], AND MICHAEL I. PETAEV[8]

[1]Hawai‘i Institute of Geophysics and Planetology, School of Ocean and Earth Science and Technology, University of Hawai‘i at Manoa, Honolulu, HI 96822, USA

[2]Lawrence Livermore National Laboratory, Livermore, CA 94451, USA

[3]Department of Earth and Planetary Sciences, Tokyo Institute of Technology, Meguro, Tokyo 152-8551, Japan

[4]Space Science Division, Ames Research Center, NASA, Moffett Field CA 94035, USA

[5]Department of Earth and Space Sciences, University of California, Los Angeles, CA 90095, USA

[6]Centre de Recherches Pétrographiques et Géochimiques, CNRS-UPR 2300, BP20, 54501 Vandoeuvre les Nancy, France

[7]Ecole Nationale Supérieure de Géologie, INPL, BP40, 54501 Vandoeuvre les Nancy, France

[8]Harvard-Smithsonian Center for Astrophysics and Department of Earth and Planetary Sciences, Harvard University, Cambridge, MA 02138, USA

*e-mail address of the correspondence author: sasha@higp.hawaii.edu

submitted to *The Astrophysical Journal*

September 27, 2004

accepted December 17, 2004

## ABSTRACT

Changes in the chemical and isotopic composition of the solar nebula with time are reflected in the properties of different constituents that are preserved in chondritic meteorites. CR carbonaceous chondrites are among the most primitive of all chondrite types and must have preserved solar nebula records largely unchanged. We have analyzed the oxygen and magnesium isotopes in a range of the CR constituents of different formation temperatures and ages, including refractory inclusions and chondrules of various types. The results provide new constraints on the time variation of the oxygen isotopic composition of the inner (<5 AU) solar nebula – the region where refractory inclusions and chondrules most likely formed. A chronology based on the decay of short-lived $^{26}$Al ($t_{1/2}$ ~ 0.73 Ma) indicates that the inner solar nebula gas was $^{16}$O-rich when refractory inclusions formed, but less than 0.8 Ma later, gas in the inner solar nebula became $^{16}$O-poor and this state persisted at least until CR chondrules formed ~1-2 Myr later. We suggest that the inner solar nebula became $^{16}$O-poor because meter-size icy bodies, which were enriched in $^{17,18}$O due to isotopic self-shielding during the ultraviolet photo dissociation of CO in the protosolar molecular cloud or protoplanetary disk, agglomerated outside the snowline, drifted rapidly towards the Sun, and evaporated at the snowline. This led to significant enrichment in $^{16}$O-depleted water, which then spread through the inner solar system. Astronomical studies of the spatial and/or temporal variations of water abundance in protoplanetary disks may clarify these processes.

## 1. INTRODUCTION

Chondritic meteorites are the oldest rocks formed in the early solar system. They consist of three major components: refractory inclusions [Ca,Al-rich inclusions (CAIs) and amoeboid olivine aggregates (AOAs)], chondrules, and fine-grained matrix (Scott & Krot 2003). CAIs are <100 μm to >1 cm-sized irregularly-shaped or rounded objects composed mostly of oxides and silicates of calcium, aluminum, titanium, and magnesium, such as corundum ($Al_2O_3$), hibonite ($CaAl_{12}O_{19}$), grossite ($CaAl_4O_7$), perovskite ($CaTiO_3$), spinel ($MgAl_2O_4$), Al,Ti-pyroxene (solid solution of $CaTi^{4+}Al_2O_6$, $CaTi^{3+}AlSiO_6$, $CaAl_2SiO_6$, and $CaMgSi_2O_6$), melilite (solid solution of $Ca_2MgSi_2O_7$ and $Ca_2Al_2SiO_7$), and anorthite ($CaAl_2Si_2O_8$). AOAs are physical aggregates of individual condensate particles − forsterite ($Mg_2SiO_4$), Fe,Ni-metal, and CAIs composed of spinel, anorthite, and Al,Ti-pyroxene. Chondrules are igneous, rounded objects, 0.01-10 mm in size, composed largely of ferromagnesian olivine ($Mg_{2-x}Fe_xSiO_4$) and pyroxene ($Mg_{1-x}Fe_xSiO_3$, where $1<x<0$), Fe,Ni-metal, and glassy or microcrystalline mesostasis. Chondrules dominated by olivine and pyroxene phenocrysts (grains crystallized from a host chondrule melt) are called porphyritic. Some few percent of all chondrules are known to be unusually rich in aluminum, and often contain relict (unmelted objects which did not crystallize from a host chondrule melt) refractory inclusions (Krot & Keil 2002; Itoh et al. 2002; Krot et al. 2001, 2002a, 2004a). Matrix material is an aggregate of mineral grains, 10 nm - 5 μm in size, that surrounds refractory inclusions and chondrules and fills in the interstices between them; in primitive chondrites, matrix is made largely of magnesian olivine and pyroxene, and amorphous ferromagnesian silicate particles (e.g., Greshake 1997).

Chondrules and refractory inclusions are believed to have formed during transient heating events in the inner (<5 AU) solar nebula (e.g., Cassen 2001; Desch & Connolly 2002; MacPherson 2003). Evaporation and condensation appear to have been the dominant processes during formation of refractory inclusions; subsequently some CAIs experienced melting to various degrees (MacPherson 2003). Multiple episodes of melting of pre-existing solids accompanied by evaporation-recondensation are believed to have been the dominant processes during chondrule formation (Desch & Connolly 2002; Scott & Krot 2003). Matrices are chemically complementary to chondrules, and may have largely experienced extensive evaporation-recondensation during chondrule formation (Kong 1999; Klerner & Palme 2000;

Scott & Krot 2003; Bland et al. 2004). Mineralogical, chemical and isotopic studies of chondritic components provide important constraints on the physical and chemical processes, physico-chemical conditions, chemical and isotopic compositions, and the lifetime of the protoplanetary disk (e.g, Krot et al. 2000a, 2002b; Russell et al. 2004; Scott & Krot 2003).

In this paper, we address the origin and evolution of oxygen isotopic composition of the inner solar nebula using oxygen and magnesium isotopic compositions of refractory inclusions and chondrules in the CR carbonaceous chondrites. CR chondrites are among the most primitive chondrite groups (Krot et al. 2002c), and largely escaped thermal metamorphism and hydrothermal alteration on their parent asteroid(s) that disturbed nebular records of both oxygen and magnesium isotopic compositions in the CV and metamorphosed CO chondrites (e.g., Krot et al. 1998; Wasson et al. 2001; Yurimoto et al. 2000; McKeegan et al. 2004). CR chondrites did experience mild, low-temperature aqueous alteration, but this appears to have not affected the oxygen and magnesium isotopic compositions of their refractory inclusions and chondrules (Marhas et al. 2000, 2001; Aléon et al. 2002; Krot et al. 2004b,c).

## 2. OXYGEN ISOTOPIC RESERVOIRS IN THE INNER SOLAR NEBULA

On a three-oxygen isotope diagram ($\delta^{18}O$ *vs.* $\delta^{17}O$, where $\delta^{17,18}O = [(^{17,18}O/^{16}O)_{sample}/(^{17,18}O/^{16}O)_{SMOW} - 1] \times 1000$); SMOW is Standard Mean Ocean Water) compositions of CAIs, AOAs, and chondrules in primitive chondrites generally plot along a line of slope ~1 and show a large range of $\Delta^{17}O$ (< -20‰ to +5‰, where $\Delta^{17}O = \delta^{18}O - 0.52 \times \delta^{17}O$) (e.g., Clayton et al. 1977; Clayton 1993; McKeegan & Leshin 2001). These *mass-independent* variations most likely resulted from mixing of $^{16}O$-rich and $^{16}O$-poor materials in the solar nebula (e.g., Clayton 1993). By contrast, *mass-dependent* variations, resulting from differences in isotopic masses during chemical and physical processes, plot along a line of slope ~1/2 (McKeegan & Leshin 2001).

The existence of an $^{16}O$-rich *gaseous* reservoir in the inner solar nebula has been inferred from the $^{16}O$-rich ($\Delta^{17}O$ < -20‰) compositions of refractory solar nebula *condensates*, such as AOAs, forsterite-rich accretionary rims around CAIs, and fine-grained, spinel-rich CAIs with Group II rare earth element (REE) patterns (volatility-fractionated REE patterns indicating condensation from a gas from which more refractory REE has already been removed; Scott &

Krot 2001; Krot et al. 2002a; Aléon et al. 2004). An $^{16}$O-poor gaseous reservoir can be inferred from most chondrules, which are consistently $^{16}$O-depleted compared to AOAs and most CAIs (e.g., Clayton 1993; Maruyama et al. 1999; Krot et al. 2004b), and from the $^{16}$O-poor compositions of aqueously formed secondary minerals in chondrites (Choi et al. 1998). Although few presolar $^{16}$O-rich solids in meteorites are known to have survived evaporation (Nittler et al. 1997, 1998), their existence has been inferred (Scott and Krot 2001) from (*i*) $^{16}$O-rich compositions of platy hibonite crystals (PLACs) and some FUN (fractionation and unknown nuclear anomalies) CAIs which are considered to be evaporative residues of the initial solar nebula dust (Fahey et al. 1987; Brigham et al. 1988; Ireland 1990; Wood 1998; Lee et al. 2001; Goswami et al. 2001), and (*ii*) from an $^{16}$O-rich magnesian chondrule, which has been interpreted as a solidified melt of this dust (Kobayashi et al. 2003).

Previous models assumed the existence of $^{16}$O-rich solids and $^{16}$O-poor gas and invoked solid-gas fractionation to produce the $^{16}$O-rich and $^{16}$O-poor solar nebula regions (Scott & Krot 2001; Krot et al. 2002b; Itoh & Yurimoto 2004). The ultimate source of the mass-independent oxygen isotope fractionation was ascribed to unknown processes, either presolar (Clayton et al. 1977; Clayton 1993; Scott & Krot 2001; Thiemens & Heidenreich 1983) or solar (Thiemens & Heidenreich 1983; Thiemens 1996). This observed mass-independent fractionation of oxygen isotopes has most recently been attributed to isotopic self-shielding during UV photolysis of CO in an *initially $^{16}$O-rich* ($\Delta^{17}$O ~ -25‰) protoplanetary disk (Kitamura & Shimizu 1983; Clayton 2002; Lyons & Young 2004) or parent molecular cloud (Yurimoto & Kuramoto 2004). According to these models, the UV photolysis preferentially dissociates C$^{17}$O and C$^{18}$O in illuminated zones of the protoplanetary disk (most likely the very uppermost, rarefied vertical layer) or the parent molecular cloud. If this process occurs in the stability field of water ice, the released atomic $^{17}$O and $^{18}$O are rapidly incorporated into water ice, while the residual CO gas becomes enriched in $^{16}$O (Yurimoto & Kuramoto 2004). Subsequent enhancement of $^{16}$O-depleted water ice relative to $^{16}$O-enriched CO gas in the midplane of the protoplanetary disk, followed by ice evaporation, can in principle result in generation of an $^{16}$O-poor gaseous region.

The history of oxygen isotopic variations in the inner solar nebula gas may have been recorded by igneous CAIs and chondrules that appear to have experienced oxygen isotopic exchange with the surrounding gas during melting (Yu et al. 1995; Yurimoto et al. 1998; Maruyama et al. 1999; Aléon et al. 2002; Itoh & Yurimoto 2003; Krot et al. 2004b,d). The timescales of CAI and chondrule formation can be deduced from their short-lived ($^{26}$Al-$^{26}$Mg) and long-lived ($^{207}$Pb-$^{206}$Pb) isotopic systematics (MacPherson et al. 1995; Amelin et al. 2002, 2004; Russell et al. 2004). Although igneous CAIs and chondrules with heterogeneous oxygen isotopic compositions are common in previously studied CV chondrites, the nature and timing of oxygen isotopic exchange affecting CV chondrite components are controversial subjects because of ubiquitous secondary alteration (e.g., Krot et al. 1998). This motivated our study of CR chondrites.

## 3. CHRONOLOGY OF REFRACTORY INCLUSIONS AND CHONDRULES

The average $^{207}$Pb-$^{206}$Pb age of two CAIs from Efremovka (CV) is 4567.2±0.6 Ma (Amelin et al. 2002). The $^{207}$Pb-$^{206}$Pb ages of chondrules range from 4566.7±1.0 Ma for Allende (CV) to 4564.7±0.7 Ma for Acfer 059 (CR) to 4562.7±0.5 Ma for Gujba (CB, an unusual, metal-rich chondrite) (Amelin et al. 2004). These observations suggest that chondrule formation started shortly after or perhaps even contemporaneously with CAI formation, and lasted for at least 4.5±1.1 Myr. The duration of CAI formation remains unknown, but could be as short as 0.5 Myr, based on $^{26}$Al-$^{26}$Mg systematics (Bizzarro et al. 2004).

Most CAIs in primitive chondrites have an initial $^{26}$Al/$^{27}$Al ratio [($^{26}$Al/$^{27}$Al)$_0$] of $5\times10^{-5}$, commonly referred as "canonical", which represents the initial abundance of $^{26}$Al in the solar nebula (MacPherson et al. 1995; Russell et al. 2004), or just in the CAI-forming region, possibly near the proto-Sun (Shu et al. 1996).[1] The rare exceptions include PLACs, FUN CAIs, and grossite-rich and hibonite-rich CAIs from CH carbonaceous chondrites, which show no evidence for live $^{26}$Al. FUN CAIs and PLACs have been interpreted as the earliest products in the solar system, which formed prior to injection and homogenization of $^{26}$Al (Sahijpal et al. 1998). The lack of $^{26}$Al in CH CAIs remains unexplained. Because the vast majority of CAIs in primitive (unmetamorphosed) chondrites are characterized by the canonical ($^{26}$Al/$^{27}$Al)$_0$ ratio, it is reasonable to assume that igneous CAIs in primitive chondrites also formed by melting of inclusions having initially canonical ($^{26}$Al/$^{27}$Al)$_0$ ratios. In contrast, the vast majority of chondrules have much lower ($^{26}$Al/$^{27}$Al)$_0$ ratios ($\leq1.5\times10^{-5}$) (Russell et al. 1996; Kita et al. 2000;

---

[1] Galy et al. (2004) reported an ($^{26}$Al/$^{27}$Al)$_0$ ratio of $(6.78\pm0.85)\times10^{-5}$ in bulk CAIs from CV chondrites, and interpreted this ratio to represent the time of segregation of the CV proto-CAIs from the rest of the solar nebula as ~0.4 Ma before the last melting of the CV CAIs that is recorded by a canonical $^{26}$Al/$^{27}$Al ratio of $5\times10^{-5}$.

Marhas et al. 2000). High $(^{26}Al/^{27}Al)_0$ ratios have been recently reported in several magnesian chondrules from the CV chondrite Allende, based on their bulk magnesium isotope measurements (Bizzarro et al. 2004).

The difference in $(^{26}Al/^{27}Al)_0$ ratios between CAIs and most chondrules suggests an age difference of at least 1-2 Myr between the crystallization ages of CAIs and the majority of chondrules (MacPherson et al. 1995; Russell et al. 1996; Kita et al. 2000; Marhas et al. 2000; Russell et al 2004). This chronological interpretation of $^{26}Al$-$^{26}Mg$ systematics is based on the assumption that $^{26}Al$ had a stellar origin and was injected and homogenized in the solar nebula over a time scale that was short compared to its half-life ($t_{1/2}$ ~ 0.7 Ma) (e.g., MacPherson et al. 1995). We note that, if interpreted chronologically, the high $(^{26}Al/^{27}Al)_0$ ratios inferred from bulk magnesium isotope measurements of the Allende chondrules (Bizzarro et al. 2004) may date the time for the formation of chondrule precursor materials, not the time of chondrule melting; the latter can be inferred from internal aluminum-magnesium isochrons, which have not been measured in these chondrules yet.

The alternative, non-chronological interpretation of $^{26}Al$-$^{26}Mg$ systematics involves a local origin of $^{26}Al$ by energetic particle irradiation near the proto-Sun, resulting in radial heterogeneity of $^{26}Al$ distribution (e.g., Gounelle et al. 2001). Although the detection in CAIs of short-lived radionuclides that can be produced by nuclear spallation reactions and not in the stars – $^{10}Be$ ($t_{1/2}$ ~ 1.5 Ma) (McKeegan et al. 2000) and possibly $^{7}Be$ ($t_{1/2}$ ~ 52 days; Chaussidon et al. 2002, 2004) – may indicate that CAIs or their precursors experienced irradiation by energetic particles, this mechanism remains problematic for $^{26}Al$ (McKeegan et al. 2000). At the same time, a stellar origin for $^{26}Al$ is consistent with (*i*) correlated abundances of $^{26}Al$ and $^{41}Ca$ ($t_{1/2}$ ~ 0.1 Ma) in CAIs (Sahijpal et al. 1998), (*ii*) discovery of $^{60}Fe$ ($t_{1/2}$ ~ 1.5 Ma) that can be produced only by a stellar nucleosynthesis in a chondrule (Huss & Tachibana 2004), and (*iii*) lack of correlation between $^{26}Al$ and $^{10}Be$ in CAIs (Marhas et al. 2002).

Here we report oxygen and $^{26}Al$-$^{26}Mg$ isotopic systematics of refractory inclusions and chondrules in CR chondrites that constrain the temporal history of $^{16}O$-rich and $^{16}O$-poor gaseous reservoirs in the inner solar nebula. We also describe an astrophysical mechanism by which the $^{16}O$-poor gas can be delivered to the chondrule-forming region on the appropriate timescale and in the appropriate abundance. Our observations indicate that the $^{16}O$-poor gaseous reservoir may have appeared rather early and persisted for at least 2 Myr, consistent with models of Lyons and Young (2004), Yurimoto and Kuramoto (2004), and Cuzzi and Zahnle (2004). However, more studies need to be done to refine the timing of this transition.

## 4. SAMPLES AND ANALYTICAL METHODS

This section describes our methodology in some detail; readers uninterested in the specifics of the measurements may skip to section 5. Twenty-nine polished sections of 16 CR chondrites (Acfer 087, Acfer 139, EET87730, EET87747, EET87770, EET92041, EET92042, EET92147, El Djouf 001, GRA95229, MAC87320, MET00426, PCA91082, Renazzo, Temple Bay, and QUE99177) were studied using optical microscopy, backscattered electron (BSE) imaging, X-ray elemental mapping, and electron probe microanalysis. X-ray elemental maps with a resolution of 2-5 µm/pixel were acquired with five spectrometers of the Cameca microprobe SX-50 operating at 15 kV accelerating voltage, 50-100 nA beam current and ~1-2 µm beam size. The Mg, Ca, and Al X-ray images were combined by using a RGB-color scheme and ENVI (Environment for Visualizing Images) software to obtain false color maps to identify all refractory inclusions and aluminum-rich chondrules suitable for oxygen and magnesium isotopic studies. BSE images were obtained with the Zeiss DSM-962, JEOL JSM-5900LV, and LEO-1430VP scanning electron microscopes using a 15-20 kV accelerating voltage and 1-2 nA beam current. Electron probe microanalyses were performed with a Cameca SX-50 electron microprobe using a 15 kV accelerating voltage, 10-20 nA beam current, beam size of ~1-2 µm and wavelength dispersive X-ray spectroscopy. The mineralogy and petrology of CAIs, AOAs and aluminum-rich chondrules in CR chondrites are described in detail in two companion papers (Aléon et al. 2002; Krot & Keil 2002) and are not discussed here.

Oxygen isotopic compositions in chondrules were measured *in situ* with the Cameca ims-1270 ion microprobe operated in multicollection mode. $^{16}O$ and $^{18}O$ were measured using Faraday cups; $^{17}O$ was measured using the axial electron multiplier. A $Cs^+$ primary beam of 10 nA was used to produce ion probe sputter pits approximately 25-30 µm in diameter. With such conditions, the count rate was $\sim 2\times10^6$ counts per second for $^{18}O$. Corrections for instrumental mass fractionation (IMF), counting statistics, and uncertainty in standard compositions were applied. The IMF was corrected using terrestrial standards: olivine, clinopyroxene, orthopyroxene, MORB glass, adularia, spinel, and quartz. Under the analytical conditions employed, the total precision (2‰ ) of individual oxygen isotopic analyses is better than 1.5‰ for both $\delta^{18}O$ and $\delta^{17}O$. Oxygen isotopic compositions in CAIs and AOAs were measured *in situ* with the UCLA Cameca ims-1270 using the operating conditions and procedures described in (Aléon et al. 2002). Following oxygen isotopic measurements, each refractory inclusion and chondrule analyzed was re-examined in backscattered electron and secondary electron images using a JEOL 5900LV scanning electron microscope to verify the locations of the sputtered craters and mineralogy of the phases analyzed.

Magnesium isotope compositions were measured *in situ* using PANURGE, a modified Cameca IMS-3f ion microprobe at Lawrence Livermore National Laboratory using the operating conditions and procedures described in (Hutcheon et al. 1987). The Mg isotope ratios were corrected for both instrumental and intrinsic fractionation assuming the standard ratios of $^{25}Mg/^{24}Mg = 0.12663$ and $^{26}Mg/^{24}Mg = 0.13932$ (Catanzaro et al. 1966). The corrected ratios $(^{26}Mg/^{24}Mg)_C$ were used to calculate $\delta^{26}Mg = [(^{26}Mg/^{24}Mg)_C/0.13932 - 1]\times1000$.

## 5. RESULTS

Oxygen isotopic compositions of the CR chondrules plot along the line of slope ~1 (Fig. 1a,b). The aluminum-rich chondrules with relict CAIs (Fig. 2a,b) have heterogeneous oxygen isotopic compositions, with the relict spinel and anorthite being $^{16}O$-enriched ($\Delta^{17}O$ = -15‰ to -10‰) compared to the chondrule phenocrysts and mesostasis ($\Delta^{17}O$ = -7‰ to -5‰). The aluminum-rich chondrules without relict CAIs (Fig. 2c), as well as magnesian porphyritic olivine and porphyritic olivine-pyroxene (Type I) chondrules, are isotopically uniform (within ±2‰ in $\Delta^{17}O$) and have similarly $^{16}O$-poor compositions ($\Delta^{17}O$ ranges from -5‰ to -1‰). Because the aluminum-rich chondrules and Type I chondrules formed by melting of various proportions of aluminum-rich and ferromagnesian precursors (Krot & Keil 2002; Krot et al. 2001, 2002a, 2004a) which had different oxygen isotopic compositions (Krot et al. 2004b), we infer that both chondrule types experienced melting and oxygen isotopic exchange in the same $^{16}O$-poor gaseous reservoir.

AOAs and most CAIs in CR chondrites are $^{16}O$-rich ($\Delta^{17}O$ ~ -20‰), while CAIs which are clearly igneous in origin[2] (compact Type A, Type B, and Type C; Fig. 3) are $^{16}O$-depleted relative to AOAs to various degrees (Fig. 1c,d). Three compact Type A CAIs and a Type B CAI (Fig. 3a-c) are only slightly less $^{16}$O-rich ($\Delta^{17}$O ~ -18‰ to -15‰) than AOAs, and have the canonical ($^{26}$Al/$^{27}$Al)$_0$ ratio of ~5×10$^{-5}$ (Fig. 4a,b). We infer that these CAIs experienced early melting in a slightly $^{16}$O-depleted gaseous reservoir.

[2] There are three major mineralogical types of igneous CAIs: (*i*) compact Type A composed of melilite with very little pyroxene, (*ii*) Type B made of melilite, spinel, pyroxene, and anorthite; and (*iii*) anorthite-rich, Type C CAIs (see MacPherson et al. 1988 for a review).

All three Type C CAIs studied – El Djouf 001 #10, MET00426 #1, and MET00426 #2 (Fig. 3d-f) – are depleted in $^{16}$O to levels comparable to those observed in the CR chondrules (Fig. 1). We infer that these CAIs experienced melting and oxygen isotopic exchange in the presence of $^{16}$O-poor gas. El Djouf 001 #10 shows striking oxygen isotopic heterogeneity, with spinel being $^{16}$O-enriched relative to anorthite and Al,Ti-pyroxene, indicating incomplete melting and oxygen isotopic exchange. MET00426 #1 and MET00426 #2 are isotopically uniform. El Djouf 001 #10 and MET00426 #2 have low ($^{26}$Al/$^{27}$Al)$_0$ ratios of (0.8±1.8)×10$^{-6}$ and <6.3×10$^{-6}$, respectively (Fig. 4d,e), similar to those in the aluminum-rich chondrules from CR chondrites (Fig. 4f; Marhas et al. 2000), suggesting a late-stage melting, > 2 Myr after formation of CAIs with the canonical ($^{26}$Al/$^{27}$Al)$_0$ ratio, possibly contemporaneously with the CR chondrules (Amelin et al. 2002). MET00426 #1 has a high ($^{26}$Al/$^{27}$Al)$_0$ ratio of (4.0±1.8)×10$^{-5}$ (Fig. 4c), suggesting melting and oxygen isotopic exchange in an $^{16}$O-poor nebula region within 0.8 Myr of CAIs with a canonical ($^{26}$Al/$^{27}$Al)$_0$ ratio forming in an $^{16}$O-rich nebular region. Below we discuss possible mechanisms and timescales for the early variation of oxygen isotopes in the inner solar nebula – the region where CAIs and chondrules most likely formed (Cassen 2001; Desch & Connolly 2002; Cuzzi et al. 2003).

## 6. $^{16}$O-RICH AND $^{16}$O-POOR GASEOUS REGIONS IN THE SOLAR NEBULA

Our data for chondrules and refractory inclusions in CR chondrites clearly indicate that the oldest, and highest-temperature, condensates (CAIs and AOAs) are the richest in $^{16}$O, and that nebula condensates interacted with a successively $^{16}$O-poorer gas as time progressed. Our observations have begun to set timescales for this transition. Below, we describe the process which we suggest was responsible for this transition, comparing its predicted timescales to our observations, and then make some comments on the initial $^{16}$O-rich state.

### 6.1. Late-stage, $^{16}$O-poor environment

According to our model and those of Clayton (2002), Lyons and Young (2004), and Yurimoto and Kuramoto (2004), initially, the nebular gas was $^{16}$O-rich. During this time $^{16}$O-rich AOAs and primitive CAIs with canonical ($^{26}$Al/$^{27}$Al)$_0$ ratio formed. The formation at a later time of $^{16}$O-*depleted* chondrules and igneous CAIs requires both the *formation* and the subsequent *transfer* of a $^{16}$O-depleted carrier (presumably $H_2O$) to the inner solar nebula separately from the complementary $^{16}$O-enriched carrier (presumably CO or primordial silicates) on a timescale that matches our observations. In the protoplanetary disk, outside 5 AU from the Sun (the inferred location of the condensation/evaporation front of water, called the "snowline"), $^{17,18}$O-enriched water ice, resulting from CO self-shielding in the molecular cloud or at high altitudes in the disk, aggregates into large particles which settle down to the disk mid-plane, causing $H_2O$/CO fractionation and enrichment of this region in heavy oxygen (Yurimoto & Kuramoto 2004). If self-shielding only occurred in the protoplanetary disk, it would have taken approximately 1 Myr to achieve the level of $^{17,18}$O-enrichment in water ice observed in aqueously formed chondritic minerals (Lyons & Young 2004). This transformation timescale greatly exceeds those for either vertical mixing, or particle growth and settling to the mid-plane (Cuzzi & Weidenschilling 2004). If CO self-shielding took place in the parent molecular cloud and the solar nebula initially contained $^{16}$O-depleted $H_2O$ and $^{16}$O-enriched CO, the transformation timescale can be significantly shorter (Yurimoto & Kuramoto 2004). Here we note that vigorous transfer of this $^{17,18}$O-enriched, outer nebula water to the inner nebula (Yurimoto & Kuramoto 2004; Lyons & Young 2004) is a natural outcome of the evaporation front process described in Cuzzi et al. (2003) and Cuzzi and Zahnle (2004). We also note implications for astronomical observations.

In the past, it has often been asserted that meter-sized particles drift radially inwards at rates which are so large that they are "lost into the sun" on short timescales. However, drifting particles would evaporate before – sometimes long before - they got that far; the full implications of this process have not been previously appreciated. A systematic study of "evaporation fronts" was first done by Cuzzi et al. (2003), in the context of CAI formation and evolution near the silicate evaporation front; a more recent study by Cuzzi and Zahnle (2004) focused on the water evaporation front. In these papers, it was shown how a small meter-sized fraction of the total solids could drift so rapidly that it dominates the net inward mass flux of all other solids. It was suggested in Cuzzi et al. (2003) and Cuzzi and Zahnle (2004), and discussed in more detail in Cuzzi and Weidenschilling (2004), that growth to meter-size is not difficult in either turbulent or nonturbulent nebulae, but that further growth can be frustrated by even a small amount of turbulence. Essentially, relative velocities between particles smaller than a meter are small enough that growth of crushable, but cohesive, particles by simple sticking is quite plausible[3]. Thus, a distribution extending from micron-sized to meter-sized particles might form very quickly and persist for a long time. In such a distribution, which has equal mass per decade radius, the mass fraction in meter-sized particles is on the order of 0.1. Weidenschilling (1997, 2000) finds such distributions quite generally at 1, 30, and 90 AU, while after different times; at 30 AU, this distribution arises before $10^5$ years have elapsed.

Water ice particles evaporate quickly, fairly close to the evaporation front (Cuzzi & Zahnle 2004; Supulver & Lin 2000). Cyr et al. (1998) modeled the evaporation of drifting water ice boulders, but the extensive radial drifts they present are not consistent with plausible nebula thermal structure, for reasons which remain unclear (Cuzzi & Zahnle 2004). Once the water has evaporated just inside the evaporation front, and become mingled with the nebular gas, it is trapped to the slow radial evolution of the nebula, and leaves the region of the evaporation front (as vapor) much slower than it arrives from outside (as solids). The abundance of water in the vapor phase thus grows significantly until its loss by inward nebular advection, and diffusion down inward and outward concentration gradients can balance the large incoming flux in drifting meter-sized particles (cf. also discussion in Yurimoto & Kuramoto 2004).

The degree of enrichment $E_o = C/Co$, where $Co$ and $C$ are respectively the cosmic and the locally enhanced abundance of water relative to hydrogen, is determined by the nebular viscosity parameter $\alpha$ and the mass fraction $f_L$ of rapidly drifting particles: $E_o \approx 2f_L/3\alpha$. Enhancements in $H_2O$ of $E_o$ = 10-100 are calculated under assumptions that the nebula has $\alpha = 10^{-2}$ to $10^{-3}$ (smaller values give larger enrichments) and that meter-sized particles constitute about $f_L = 0.1$ of the local solid mass; this in turn relies on modeling of incremental growth with erosion and destruction, which indicates that particles can indeed grow to roughly meter-size in times of 100-1000 years even if the nebula is weakly turbulent (Weidenschilling 1997; Cuzzi et al. 2003; Cuzzi & Weidenschilling 2004; Weidenschilling & Cuzzi 2004). The specific value of $E_o$, and the specific $^{16}$O-depletion of the $H_2O$, are coupled and remain to be constrained. This can be done by measurements of oxygen isotopic composition of water in cometary ices, or by astronomical observations of the relative abundance of water in the vapor form at different locations in protoplanetary disks of different ages and accretion rates (thermal structures).

[3] This conclusion is also reached by recalculating particle size limits following Weidenschilling (1988) under lower nebula turbulent intensity than the $\alpha = 10^{-2}$ assumed therein.

Analytical solutions of the equations governing this drift-diffusion situation show three different regimes (Cuzzi & Zahnle 2004). At very small times, a plume of enhanced water appears just inside the evaporation front (Figure 5; regime 1). It takes a certain amount of time for this plume to spread into the inner solar system. Spreading occurs due both to the steady, but slow, inward drift of the nebula gas, and also due to diffusion down the inward concentration gradient of regime 1. The timescales for these processes are comparable: $t_{ss}$ = 0.5-5 Myr, if nebula evolution is by turbulent viscosity (Cuzzi & Zahnle 2004). After a time $t_{ss}$, the solutions indicate a uniformly enhanced abundance everywhere inside the evaporation front (Figure 5; regime 2). There may be a third regime, initially described by Stevenson and Lunine (1988), in which an immobile sink or "cold finger" forms outside the evaporation front and depletes the inner nebula in volatiles (regime 3 of Cuzzi & Zahnle 2004; not discussed here). These regimes might have other implications for meteoritics than discussed here (*cf.* Cuzzi & Zahnle 2004).

Since chondrule formation involved melting of solids during multiple heating episodes (Desch & Connolly 2002), and appears to have persisted for at least 2 Myr after the formation of pristine CAIs (Russell et al. 1996; Kita et al. 2000; Marhas et al. 2000; Amelin et al. 2002, 2004), these models are consistent with most chondrules exchanging oxygen with an $^{16}$O-poor nebular gas in regime 2. Chondrules and fine-grained matrices are chemically complementary, suggesting that they formed in the same location (Kong 1999; Klerner & Palme 2000; Bland et al. 2004). Because most of the fine-grained matrix materials are expected to be vaporized and recondensed during chondrule-forming events (Desch & Connolly 2002), they would also have exchanged their oxygen with an $^{16}$O-poor nebular gas. This suggestion is consistent with the relatively $^{16}$O-poor compositions of chondrules and matrix materials (this paper; also Clayton et al. 1977; Clayton 1993; Maruyama et al. 1999; Krot et al. 2004b; Jones et al. 2004; Nagashima et al. 2004).

### 6.2. Locally early- and late-stage, $^{16}$O-rich environment

If the initial oxygen isotopic composition of dust and gas in the presolar molecular cloud was $^{16}$O-rich (GENESIS results will hopefully clarify this issue) as suggested by (Clayton 2002; Yurimoto & Kuramoto 2004) and if $^{16}$O-rich presolar dust was evaporated in more than their nominal cosmic abundance in the inner nebula, the ensuing gas would become $^{16}$O-rich *regardless* of the isotopic composition of the nebular gas (Cassen 2001; Scott & Krot 2001; Kobayashi et al. 2003; Itoh & Yurimoto 2003; Cuzzi et al. 2003). An enrichment of the nebular source region of CAIs and AOAs in ice-free silicates by a factor of 10-100 relative to the solar composition is also necessary to account for the observed range of FeO contents in the $^{16}$O-rich olivine and low-Ca pyroxene condensates in AOAs (Krot et al. 2004c,e; Nagashima et al. 2004). In a process analogous to the water evaporation front discussed above, silicate material could have been comparably enhanced inside the silicate evaporation boundary (~1375 K for forsterite at total pressure of $10^{-4}$ bar; Cuzzi et al. 2003). Models of evolving nebulae by, for instance, Bell et al. (1997), Stepinski (1998) and Cassen (2001), and widely scattered observational data (Calvet et al. 2000) tend to show mass accretion rates in the range of $10^{-6}$ to $10^{-7}$ solar masses/yr for, typically, $10^4$ - $10^5$ years. For these accretion rates, midplane temperatures would much higher than the photospheric temperatures directly observed and could exceed the evaporation temperature of forsterite in the terrestrial planet region (Woolum & Cassen (1999).

However, such local $^{16}$O-rich environments could not have formed after the inner nebula became too cool to evaporate silicates. In addition, $^{16}$O-rich (i.e., primordial) silicates could be difficult to preserve in the later epoch of the solar nebula due to their cumulative thermal processing under conditions known to be increasingly $^{16}$O-poor. Moreover, the relatively high abundances of moderately volatile elements (Mn, Cr, Na) in typical chondrules suggest that chondrules formed at lower ambient temperatures than CAIs and AOAs, i.e., below the condensation temperature of forsterite (Krot et al. 2002c). Thus it is very unlikely that chondrules could have formed in a locally generated, late-stage, $^{16}$O-rich environment. Indeed, uniformly $^{16}$O-rich chondrules are extremely rare[4]; only one uniformly $^{16}$O-rich chondrule has been reported so far (Kobayashi et al. 2003).

[4] Jones et al. (2004) described two Type I chondrules in a CV chondrite with olivine grains displaying a wide range of oxygen isotopic composition ($\Delta^{17}$O from 0 up to -25‰). The chondrules, however, have $^{16}$O-poor bulk compositions ($\Delta^{17}$O ~ -5‰), suggesting that they formed by melting of $^{16}$O-rich precursors in an $^{16}$O-poor gaseous reservoir and experienced incomplete oxygen isotopic exchange.

## 7. SUMMARY

In this paper we have shown that refractory inclusions in primitive CR chondrites which solidified (either as condensates or from melts) early, and generally at high temperatures, equilibrated with a gas which was richer in $^{16}O$ than later forming solid particles. We have suggested a mechanism by which $^{16}O$-poor water, accreted as solids in the outer nebula from UV-photodissociated gas, was delivered to the inner nebula in significant amounts, and on fairly short timescales (0.5-5 Myr). We found one igneous CAI which shows an $^{26}Al$-$^{26}Mg$ relative age of $\leq$ 0.8 Myr after pristine CAIs with the canonical $(^{26}Al/^{27}Al)_0$ ratio and still appears to have formed in an $^{16}O$-*poor* environment. The similarly $^{16}O$-poor compositions of chondrules in CR chondrites suggest that this state persisted at least until the CR chondrules formed, i.e. ~1-2 Myr later. Additional observations of this type will refine the timescale on which the transition between the $^{16}O$-rich and $^{16}O$-poor environments occurred.

Astronomical observations also have the potential to address these questions in several ways. The midplane temperatures in accreting disks are of great importance; thus it is critical to distinguish between photospheric temperatures and midplane temperatures. Woolum and Cassen (1999) have taken some first steps in this direction. Najita et al. (2003) have shown that actively accreting protoplanetary disks appear to have abundant gas phase CO emission which extends well out into the terrestrial planet region; Carr et al. (2004) have found that, in some cases, disks or disk regions showing abundant CO emission show little water emission. These effects might be explained by differences in the excitation processes from place to place or time to time (Najita et al. 2003), but perhaps future work can begin to separate out excitation effects from abundance variations and provide evidence for or against zones of differentially enhanced (or depleted) water and/or CO, or other molecules, in the vapor phase at different locations in protoplanetary disks of different ages.

Finally, we note that if chondrule formation initiated planetesimal accretion in the inner solar system (Scott 2003), the inner solar system planetesimals and planets should be $^{16}O$-poor, like chondrules; this is consistent with the existing data on Earth and meteorites from asteroids and Mars (e.g., Clayton 1993; McKeegan & Leshin 2001). On the surface this seems a rather unremarkable conclusion, but it might serve to rule out formation of the planets from earlier-formed planetesimals which have only a thin surface veneer of chondritic material.

**ACKNOWLEDGEMENTS:**

This work was supported by NASA grants NAG5-10610 (A. N. Krot, P.I.), NAG5-11591 (K. Keil, P.I.), UPN 344-37-22-03 (J. N. Cuzzi, P.I.), W-19,894 (I. D. Hutcheon, P.I.), NAG5-4704 (K. D. McKeegan, P.I.), and Monkasho grants (H. Yurimoto, P.I.). This work was performed under the auspices of the U. S. Department of Energy by the University of California, Lawrence Livermore National Laboratory under Contract No. W-7405-Eng-48. This is Hawai‘i Institute of Geophysics and Planetology publication No. XXX and School of Ocean and Earth Science and Technology publication No. YYY.

## FIGURE CAPTIONS

Fig. 1. Oxygen isotopic compositions of chondrules and refractory inclusions in CR carbonaceous chondrites. Oxygen isotopic compositions of the refractory inclusions and chondrules in Figures 1a and 1c are plotted as $\delta^{17}O$ *vs.* $\delta^{18}O$ ($\delta^{17,18}O$ = $[(^{17,18}O/^{16}O)_{sample}/(^{17,18}O/^{16}O)_{SMOW} - 1] \times 1000$); SMOW is Standard Mean Ocean Water; 2σ error bars, not shown for clarity, are ~ 2‰ and ~1‰ for $\delta^{17}O$ and $\delta^{18}O$, respectively]; the data follow mass-independent fractionation [Carbonaceous Chondrite Anhydrous Mineral (CCAM)] line. In Figures 1b and 1d, the same data are plotted in the form of deviation from Terrestrial Fractionation line, as $\Delta^{17}O$ ($\Delta^{17}O = \delta^{18}O - 0.52 \times \delta^{17}O$); error bars are 2σ. Each column represents data for a single chondrule or a refractory inclusion (chondrule and refractory inclusion numbers are indicated) and shows variations in oxygen isotopic composition within an individual object. Aluminum-rich chondrules and magnesium-rich (Type I) chondrules are $^{16}O$-depleted relative to amoeboid olivine aggregates (AOAs) and most Ca,Al-rich inclusions (CAIs). Aluminum-rich chondrules with relict CAIs (see Fig. 2a,b) are $^{16}O$-enriched compared to those without relict CAIs (see Fig. 2c) and Type I chondrules. Clearly igneous [Compact Type A (CTA), Type B, and Type C] CAIs are $^{16}O$-depleted compared to AOAs and non-igneous (?) CAIs. Three Type C CAIs (shown in Fig. 3d-f) are $^{16}O$-depleted to a level observed in the CR chondrules. Aluminum-rich chondrules with relict CAIs show oxygen isotopic heterogeneity with spinel being $^{16}O$-enriched relative to anorthite and chondrule phenocrysts. Aluminum-rich chondrules without relict CAIs and Type I chondrules are isotopically uniform. The only exception is a Type I chondrule #11 from MAC87320 that contains an $^{16}O$-enriched relict olivine grain. an = anorthite or anorthitic plagioclase; cpx = high-Ca pyroxene; grs = grossite; hib = hibonite; mel = melilite; mes = glassy or microcrystalline mesostasis; ol = forsteritic olivine; opx = low-Ca pyroxene; px = Al,Ti-pyroxene; sp = spinel. Igneous CAIs: CTA #1 = GRA95229 #1; CTA #2 = GRA95229 #2; CTA #3 = GRA95229 #3; Type B #22 = GRA95229 #22; Type C #1 = MET00426 #1; Type C #2 = MET00426 #2; Type C #10 = El Djouf 001 #10.

Fig. 2. Combined elemental maps in Mg (red), Ca (green), and Al Kα (blue) X-rays of the aluminum-rich chondrules #1 from EET92147 (a), #1 from El Djouf 001 (b), and #1 from MAC87320 (c). The chondrules consist of forsteritic olivine (ol), low-Ca pyroxene (px), high-Ca pyroxene (cpx), anorthitic plagioclase (pl), Fe,Ni-metal (met), and fine-grained mesostasis

composed of silica, high-Ca pyroxene, and anorthitic plagioclase. Chondrule EET92147 #1 contains an incompletely melted, relict CAI composed largely of spinel (sp), forsteritic olivine, and anorthitic plagioclase. Chondrule El Djouf 001 #1 experienced more complete melting and contains only rare relict spinel grains; no relict grains are observed in the most extensively meleted chondrule MAC87320 #1. Oxygen isotopic compositions of the chondrules EET92147 #1 and MAC87320 #1 are shown in Figure 1b.

Fig. 3. Backscattered electron images of the igneous, $^{16}O$-rich (a-c) and $^{16}O$-poor (d-f) CAIs from the CR carbonaceous chondrites. Type B CAI GRA95229 #22 (a), compact Type A (CTA) CAI GRA95229 #2 (b), CTA CAI GRA95229 #3 (c), and Type C CAIs El Djouf 001 #10 (d), MET00426 #2 (e), and MET00426 #1 (f). an = anorthite; di = Al,Ti-diopside; mel = melilite; mes = mesostasis; met = Fe,Ni-metal; ol = forsteritic olivine; pv = perovskite; sp = spinel. Oxygen isotopic compositions of these CAIs are shown in Figure 1d.

Fig. 4. The aluminum-magnesium evolutionary diagrams of (a) the $^{16}O$-enriched compact Type A CAI GRA95229 #2, (b) the $^{16}O$-enriched Type B CAI GRA95229 #22, (c-e) the $^{16}O$-depleted/poor Type C CAIs (c) MET00426 #1, (d) MET00426 #2, (e) El Djouf 001 #10, and (f) the $^{16}O$-poor Al-rich chondrule with a relict CAI EET92147 #1. The $(^{26}Al/^{27}Al)_0$ ratios in the $^{16}O$-enriched CAIs are close to the canonical ratio of $5\times10^{-5}$, suggesting an early melting. The high $(^{26}Al/^{27}Al)_0$ ratio in the $^{16}O$-poor Type CAI, MET00426 #1 suggests that its melting and oxygen isotopic exchange occurred not more than 0.8 Myr after formation of CAIs with the canonical $(^{26}Al/^{27}Al)_0$ ratio. The $(^{26}Al/^{27}Al)_0$ ratios in the $^{16}O$-poor CAIs El Djouf 001 #10 and MET00426 #2, and the Al-rich chondrule with relict CAI are low, suggesting they experienced melting and oxygen isotopic exchange > 2 Myr later.

Fig. 5. Schematic of the radial and temporal variation of enhancement $E_o = C/C_o$ for water with an evaporation boundary at $R_{ev}$ = 5 AU. Regime 1 represents the transition situation, where the inner solar nebula retains $E_o$ = 1 for up to $t_{ss} = 40/\alpha$ orbit periods. Regime 2 is the steady state situation for times longer than $t_{ss}$, where the inner solar nebula is enriched in water (as vapor) by a factor of 20 in this example. It takes between 0.5-5 Myr (depending on nebular properties) for the enhancement water plume at 5 AU to spread throughout the inner solar system into its steady state configuration (after Cuzzi & Zahnle, 2004).

Fig. 1.

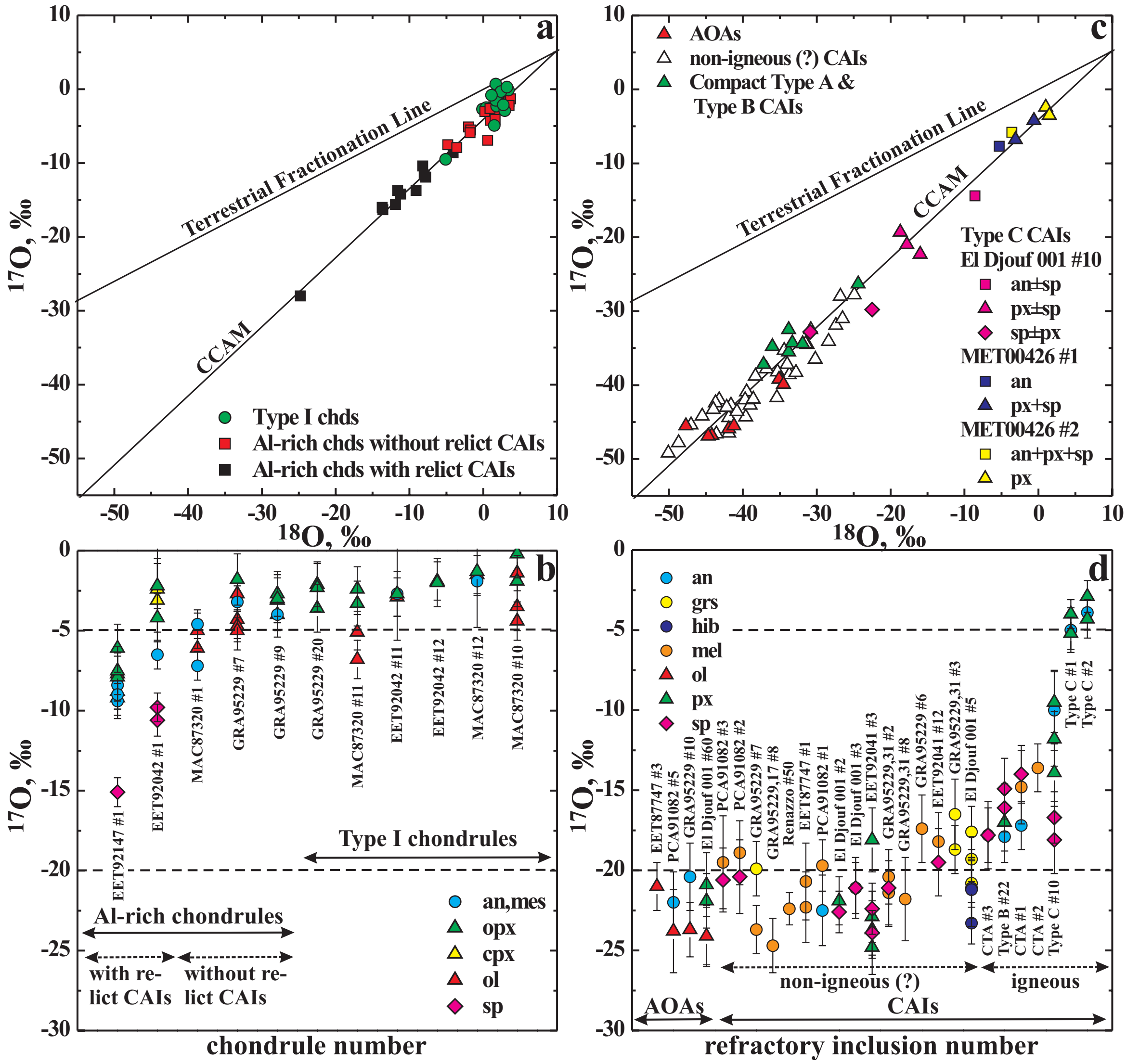

a
b
c
d
Terrestrial Fractionation Line
CCAM
Type I chds
Al-rich chds without relict CAIs
Al-rich chds with relict CAIs
AOAs
non-igneous (?) CAIs
Compact Type A & Type B CAIs
Type C CAIs
El Djouf 001 #10
an±sp
px±sp
sp±px
MET00426 #1
an
px+sp
MET00426 #2
an+px+sp
px
17O, ‰
18O, ‰
Al-rich chondrules
with relict CAIs
without relict CAIs
Type I chondrules
an,mes
opx
cpx
ol
sp
chondrule number
an
grs
hib
mel
ol
px
sp
AOAs
non-igneous (?)
CAIs
igneous
refractory inclusion number

Fig. 2.

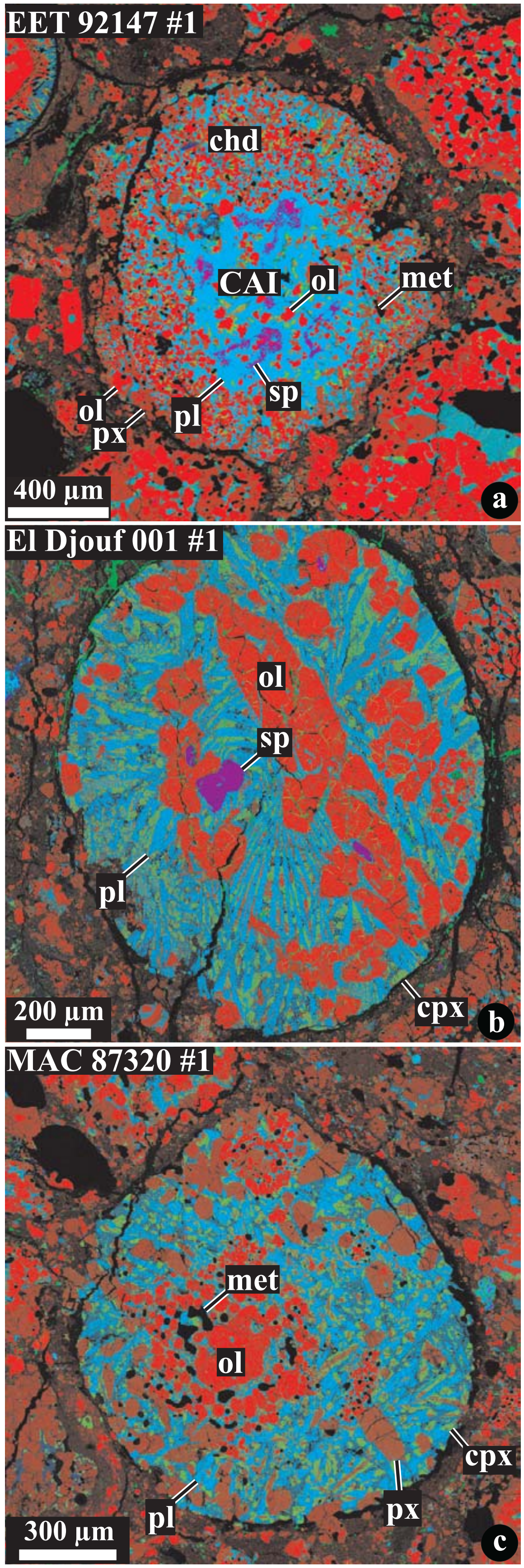
EET 92147 #1
chd
CAI
ol
met
ol
px
pl
sp
400 µm
a
El Djouf 001 #1
ol
sp
pl
cpx
200 µm
b
MAC 87320 #1
met
ol
cpx
pl
px
300 µm
c

Fig. 3.

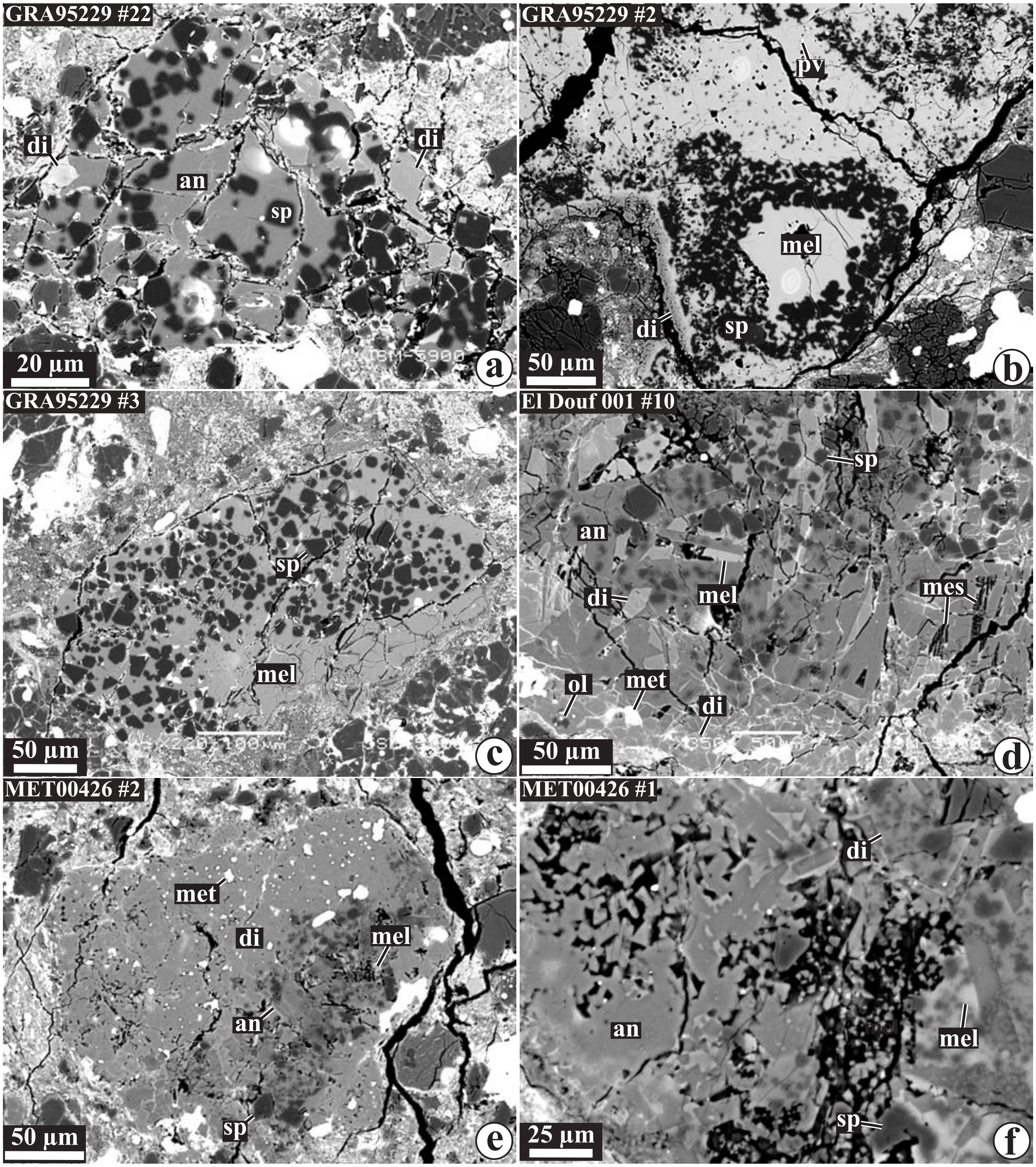
GRA95229 #22
di
di
an
sp
20 µm
a
GRA95229 #2
pv
mel
di
sp
50 µm
b
GRA95229 #3
sp
mel
50 µm
c
El Douf 001 #10
sp
an
di
mel
mes
ol
met
di
50 µm
d
MET00426 #2
met
di
mel
an
sp
50 µm
e
MET00426 #1
di
an
mel
sp
25 µm
f

Fig. 4.

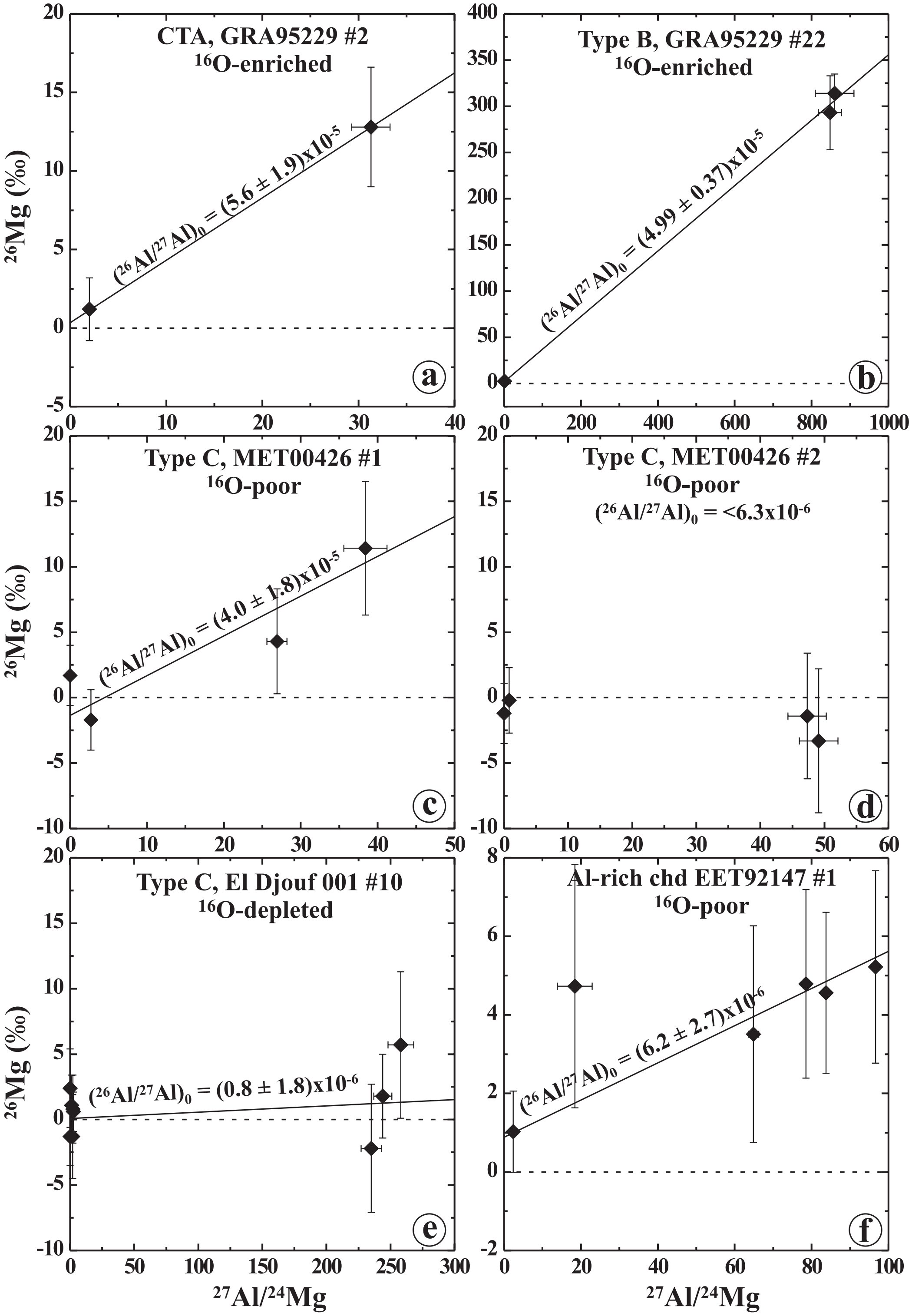
CTA, GRA95229 #2
16O-enriched
(26Al/27Al)0 = (5.6 ± 1.9)x10-5
a
Type B, GRA95229 #22
16O-enriched
(26Al/27Al)0 = (4.99 ± 0.37)x10-5
b
Type C, MET00426 #1
16O-poor
(26Al/27Al)0 = (4.0 ± 1.8)x10-5
c
Type C, MET00426 #2
16O-poor
(26Al/27Al)0 = <6.3x10-6
d
Type C, El Djouf 001 #10
16O-depleted
(26Al/27Al)0 = (0.8 ± 1.8)x10-6
e
Al-rich chd EET92147 #1
16O-poor
(26Al/27Al)0 = (6.2 ± 2.7)x10-6
f
26Mg (‰)
27Al/24Mg

Fig. 5.

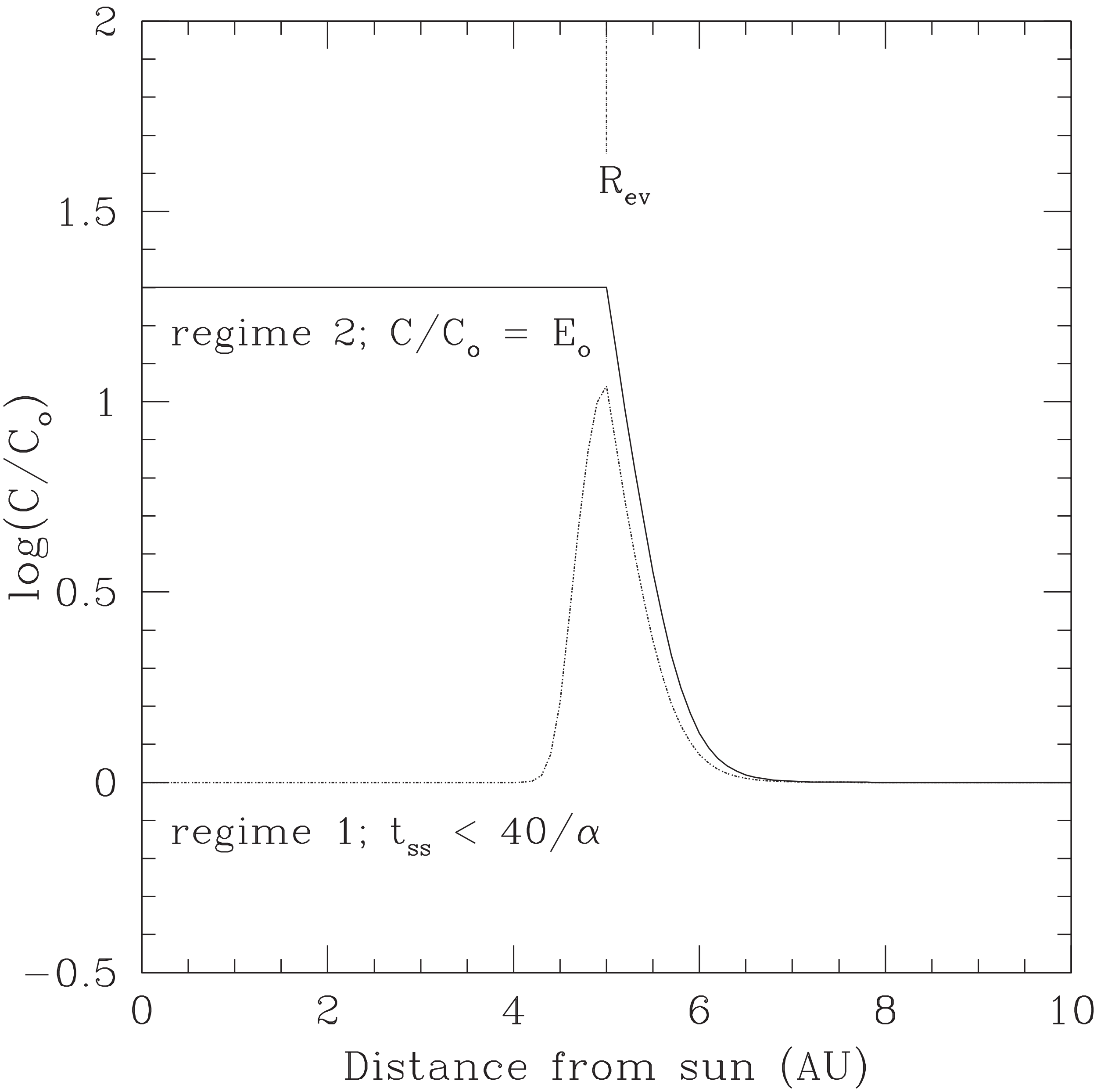

2
1.5
1
0.5
0
−0.5
0
2
4
6
8
10
R_ev
regime 2; C/C_o = E_o
regime 1; t_ss < 40/α
log(C/C_o)
Distance from sun (AU)